\documentclass[aps,prl,groupedaddress,superscriptaddress,twocolumn]{revtex4}

\usepackage{graphicx}
\usepackage{epstopdf}
\usepackage{amssymb}
\usepackage{color}
\usepackage{amsmath}

\begin{document}

\title{Threshold voltage and space charge in organic transistors}

\author{I. Guti\'{e}rrez Lezama}
\author{A. F. Morpurgo}
\affiliation{DPMC and GAP, Universit\'{e} de Gen\'{e}ve, 24 quai Ernest Ansermet, CH1211 Geneva, Switzerland}

\date{\today}

\begin{abstract}
We investigate  rubrene single-crystal field-effect transistors,
whose stability and reproducibility are sufficient to measure
systematically the shift in threshold voltage as a function of
channel length and source-drain voltage. The shift is due to
space-charge transferred from the contacts, and can be modeled
quantitatively without free fitting parameters, using Poisson's
equation, and by assuming that the density of states in rubrene is
that of a conventional inorganic semiconductor. Our results
demonstrate the consistency, at the quantitative level, of a variety
of recent experiments on rubrene crystals, and show how the use of
FET measurements can enable the determination of microscopic
parameters (e.g., the effective mass of charge carriers).
\end{abstract}

\maketitle

Organic single-crystal field-effect transistors (FETs) are opening
new possibilities for the detailed investigation of the intrinsic
electronic properties of organic semiconductors and of their
interfaces \cite{deBoer04, Gershenson06}. Transistors where a
single-crystal was suspended on top of a gate electrode, have led to
the observation of intrinsic transport properties, such as mobility
anisotropy \cite{Sundar04} and metallic-like temperature dependence
\cite{Podzorov04}. Current work is aiming at the systematic study of
microscopic electronic processes in these systems. Examples are the
study of polaronic effects at the interface between organic crystals
and highly polarizable dielectrics \cite{Stassen04,Hulea06,
Fratini08}, the analysis of band-like transport at interfaces with
low-$\epsilon$ materials \cite{Podzorov05, Pernstich08}, and the
detailed investigation of electronic transport at metal/organic
interfaces \cite{Molinari08}. In most cases, a quantitative analysis
of the data in terms of well-defined microscopic models has been
possible, but the consistency of results obtained in different
experiments remains to be verified.

Virtually all experiments on single crystal FETs have focused on
transport through a well-formed conducting channel, i.e. in the
regime when the gate voltage is biased well above the threshold
voltage $V_T$. Here, we use rubrene single-crystal FETs for a
systematic experimental investigation of the behavior of the
threshold voltage itself. Specifically, we have measured the
electrical characteristics of short channel transistors as a
function of channel length $L$, and extracted the dependence of
$V_T$ on $L$ and on source-drain bias $V_{DS}$. We find that $V_T$
systematically shifts to more positive values when $L$ is decreased
or $V_{DS}$ is increased, a behavior originating from changes in the
space-charge transferred from the contacts into the semiconductor.
We model the system using Poisson's equation, under the assumption
that the density of states in the molecular crystals has the same
functional dependence as in inorganic semiconductors, and find
excellent quantitative agreement between experimental data and
calculations, without introducing any adjustable parameter. Our
results indicate that the effective mass of carriers in the rubrene
valence band is close to the free electron mass, provide information
about the low-energy density of states, and show that the physical
picture used to interpret a variety of recent experiments
\cite{Molinari08, Li07} is internally consistent at a quantitative
level.

\begin{figure}[htbp]
\includegraphics[width=0.8\linewidth]{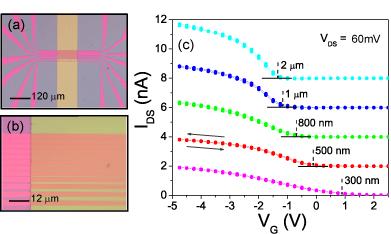}
  \caption {Panels (a) and (b) show optical microscope images of devices used in this work. Panel (c) shows the transfer characteristics of FETs with different channel lengths, measured at $V_{DS}= 60$ mV (the curves are offset vertically
  for clarity). The sub-linear $V_G$ dependence is characteristic of contact dominated devices \cite{Molinari08}. The vertical arrows point to the position of the threshold voltage in different devices. The two
  arrows with a continuous line indicate that for each data set the gate voltage is swept up and down.}
  \label{FIG. 1}
\end{figure}

Not much is currently known about the microscopic processes
determining the behavior of the threshold voltage in organic
transistors. Work done on thin-film devices has established that
this behavior can be complex, exhibiting large sample-to-sample
fluctuations and instabilities \cite{VGStress}. These phenomena have
so far prevented systematic experiments, and are posing severe
technological problems. The situation appears to be different in
single-crystal devices, with recent work showing considerably better
reproducibility \cite{Reese09}.

The fabrication of the rubrene FETs relies on the lamination of thin
($\leq$ $1$ $\mu m$-thick) single-crystals on top of a highly doped
silicon substrate (acting as a gate) covered with 500 nm SiO$_2$,
with predefined oxidized copper electrodes \cite{deBoer03}. The
devices are identical to those that we used previously for the
investigation of the contact resistance, and are therefore very well
characterized \cite{Molinari08, Molinari07}. In that work we focused
on the high-gate voltage regime with a fully formed channel; here we
analyze the behavior of the threshold voltage $V_T$.

Transport measurements were performed at room temperature, in the
dark and in vacuum ($2-7$ $10^{-7}$ mbar), using a HP4156A parameter
analyzer. Owing to the short channel length, the resistance of the
devices is dominated by the contacts, resulting in the
characteristic (sub-linear) shape \cite{Molinari08} of the transfer
curves seen in Figs. 1 and 2. Special care was taken in restricting
the applied gate voltage $V_{G}$ to the narrowest possible range
around $V_T$, to suppress possible effects of bias stress.
Accordingly, the output characteristics of our devices are
hysteresis-free, with stable and reproducible $V_{T}$ values. To
extract these values, we have used the linear extrapolation method
and fitted the linear regime of the transfer curve that follows the
onset of conduction (as illustrated in the inset of Fig. 2). The
uncertainty on the $V_T$ values extracted using this method is at
most $\pm$ 0.1 V (for the shortest devices), and usually better.

\begin{figure}[htbp]
\includegraphics[width=0.80\linewidth]{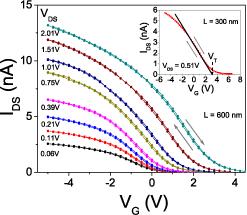}
  \caption {Transfer characteristics of a rubrene single-crystal FET with channle length $L=600$ nm, measured for
   different values of the source drain voltage $V_{DS}$. The arrows indicate that in all measurements the gate voltage has been swept up and down. The inset illustrates how the threshold voltage is
    extracted from the data. }
  \label{FIG. 2}
\end{figure}

\begin{figure}[htbp]
\includegraphics[width=0.80\linewidth]{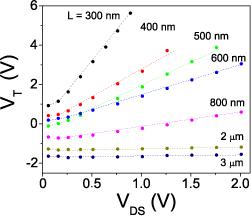}
  \caption {The dots represent $V_T$ values of short-channel single-crystal devices as a function of $V_{DS}$ measured on devices with different channel length $L$ (the dotted lines are guides for the eye). }
  \label{FIG. 3}
\end{figure}

The behavior of the threshold voltage as a function of channel
length $L$ and bias $V_{DS}$ is apparent from Figs. \ref{FIG. 1}c
and \ref{FIG. 2}.  Fig. \ref{FIG. 1}c shows the transfer
characteristics of transistors with different channel lengths,
measured at a same value of $V_{DS}$. It is apparent that $V_T$
systematically shifts to more positive values in devices with a
shorter channel. Fig. \ref{FIG. 2} shows measurements done  on an
individual transistor for increasing values of $V_{DS}$. A shift of
$V_T$ to positive values (see Fig. 3) is clearly present, which is
found to be larger for shorter devices. These trends resemble those
recently reported in Ref. \cite{Reese09}.  Fig. \ref{FIG. 3}
provides a complete overview of the behavior of $V_{T}$ in our
experiments. The different devices (i.e., sets of FETs with
different channel length fabricated on a same crystal) investigated,
exhibited identical trends and similar magnitude of the observed
effects.

The shift of $V_T$ as a function of $L$ and $V_{DS}$ originates from
the transfer of charge from the source and drain contacts into the
semiconductor. At low $V_{DS}$ (i.e., $V_{DS} \simeq kT/e$), charge
transfer occurs to align the Fermi level in the metal and in the
bulk of the semiconductor. As the contact separation is decreased,
the corresponding charge density in the (bulk) region of the
semiconductor between source and drain contact increases. At finite
$V_{DS}$, additional charge is transferred from the contacts to the
semiconductor due to capacitive coupling, similarly to what happens
in space-charge limited current measurements \cite{SCLC}. As charge
density and potential are linked by Poisson's equation, the
transferred charge induces a shift in electrostatic potential, which
needs to be compensated by a shift in the threshold voltage to
switch off conduction. In simpler terms, to switch off conduction,
the space charge transferred from the electrodes into the bulk of
the semiconductor needs to be compensated by an equal and opposite
amount of charge accumulated in the channel (i.e., by a shift in $V_T$).

\begin{figure}[htbp]
\includegraphics[width=1\linewidth]{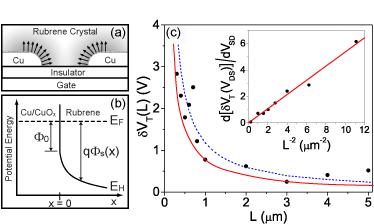}
  \caption {Panel (a) schematics of the region between the source and drain contacts of a FET device.
  The arrows indicate charge transfer from the contacts into the semiconductor; the transferred charge is represented by the shaded region.
  Panel (b) shows the potential energy band-diagram near one metal/rubrene interface.
  Panel (c) shows the comparison between the measured $\delta V_{T}(L)$ and the values calculated using Poisson's equation. The blue dotted line and the red line represent the values
calculated with $\Phi_0$ $=$ $0.13$ and $0.15$ $eV$, respectively
(corresponding to the known range of measured values for the
CuO$_x$/rubrene interface). The inset shows that $d[\delta
V_T(V_{DS})]/dV_{DS}$ scales linearly with $L^{-2}$. }
  \label{FIG. 4}
\end{figure}

At a quantitative level we consider the case of low-bias first, and
we calculate the profile of charge density transferred from the
electrodes into the bulk of the semiconductor (Fig. \ref{FIG. 4}a)
by solving Poisson equation. To this end we employ a one-dimensional
(1D) model, to describe how electrostatic potential $\Phi_s(x)$  and
charge density $\rho_s(x)$ vary in the bulk of the semiconductor,
away from an interface with a metal contact (Fig. 4b). We then have
\begin{equation}\label{Poisson}
\frac{d^2\Phi_s(x)}{dx^2}=\frac{q\rho_s(x)}{\epsilon_0\epsilon_C}=
\frac{q}{\epsilon_0\epsilon_C}M_He^{-q\Phi_s(x)/kT}.
\end{equation}
Here $M_H= 2(2\pi m_p kTh^{-2})^{3/2}$, with $m_p$ the effective
mass of holes in the "highest occupied molecular orbital" (HOMO)
band of rubrene (i.e., the valence band of the organic
semiconductor), $\epsilon_0$ is the vacuum permittivity and
$\epsilon_C$ the relative dielectric constant of rubrene. The last
equality assumes that the density of states in rubrene has the same
dependence as in a conventional one-band semiconductor, i.e.
parabolic dispersion relation with effective mass $m_p$, which
allows expressing the local density of charge carriers $\rho_s(x)$
as a function of the local potential $\Phi_s(x)$ \cite{As&Me}.
Clearly, the validity of this hypothesis needs to be checked by
comparing the results of the calculations to the experimental data.

There are two boundary conditions associated to Poisson's equation.
The first, $\Phi_s(0)$ = $\Phi_0/q$, fixes the constant in the
potential so that $q\Phi_s(x)$ corresponds to the local distance
between the Fermi level and the valence band edge in the
semiconductor ($\Phi_0$ is the energy difference between the metal
Fermi level and the valence band edge; when a large gate voltage is
applied $\Phi_0$ corresponds to the height of the Schottky barrier
present at the metal/semiconductor interface). The second is
$d\Phi_s/dx|_{x\longrightarrow \infty}=0$, which imposes that the
electric field vanishes in the bulk of the semiconductor, far away
from the interface with the metal. Under these conditions, the
solution to Poisson's equation is
\begin{equation}\label{Potential}
\ \Phi_s(x)=\frac{2kT}{q}\ln \left [e^{\Phi_0/2kT} +
x\sqrt{\frac{q^2M_H}{2kT\epsilon_0\epsilon_C}} \right]
\end{equation}
and the corresponding charge carrier density is given by
\begin{equation}\label{Density}
\ \rho_s(x)=qM_H \left [e^{\Phi_0/2kT} +
x\sqrt{\frac{q^2M_H}{2kT\epsilon_0\epsilon_C}} \right]^{-2}
\end{equation}

Note that in Eq. \ref{Poisson} we have neglected "bulk"
contributions to the charge density, e.g. due to (unintentionally
present) dopants or thermally activated charge carriers.
Accordingly, equations \ref{Potential} and \ref{Density} are valid
inside the semiconductor, only at a small distance from the
electrode surface, where the charge transferred from the contacts is
larger than the one present due to dopants, which in high-purity
crystals is $N_{D}$ $\approx$ $2$ $10^{14}$ $cm^{-3}$ \cite{Kaji09}.
This value is much smaller than what is usually found in organic
semiconductors, and hence in our devices the space charge penetrates
deeper into the rubrene crystals. Using Eq. \ref{Density}, with
$\Phi_0$ $=$ $0.13-0.15$ $eV$ \cite{Molinari08} (see discussion
below) and taking into account the overlapping of the space charge
regions coming from both contacts, it is easy to verify that the
charge transferred from the contacts is larger than $N_D$ even for
devices with $L$ $\simeq$ $1$ $\mu m$ (normally, due to the large
doping in organic semiconductors, this length is only a few tenths
of nanometers \cite{Worne08}).

To link the calculated quantities to the data, we impose that the
total amount of space charge transferred from the contacts into the
bulk region of the semiconductor between source and drain contacts
is compensated, at threshold, by an equal and opposite amount of
charge accumulated in the channel (i.e., by a shift in threshold
voltage $\delta V_T(L)$). We then get
\begin{equation}\label{LShift}
    \begin{split}
        \delta V_T(L)&=\frac{qt_C<\rho_s(L)>}{C_G} \\
        &=\frac{2qt_CM_H }{C_G} \left[e^{\Phi_0/kT}+Le^{\Phi_0/2kT}\sqrt{\frac{q^2M_H}{2kT\epsilon_0\epsilon_C}}\right]^{-1}
    \end{split}
\end{equation}
where $<\rho_s(L)>$ is the spatial average of the carrier density
present in the overlapping space-charge regions originating from the
two contacts, $C_G$ is the gate capacitance per unit area, and $t_C$
is the thickness of the rubrene crystal (measured with a
profilometer). This estimate of $\delta V_T$ is approximate, in that
our calculation does not take into account the precise geometry. For
our devices, we estimate that the error is a geometrical factor of
the order of unity (see the below discussion on the $V_{DS}$
dependence of $\delta V_T$).

Fig. \ref{FIG. 4}c shows the comparison between the predictions of
equation \ref{LShift} and the measured values $\delta V_{T}(L)$ $=$
$V_T(L)$ $-$ $<V_T>$ (where $<V_T>$ $=$ $-1.9$ $V$ is the average
threshold voltage obtained from devices with 10 $\mu m$ $\leq$ L
$\leq$ 50 $\mu m$; all measurements were performed at $V_{DS}$ $=$
$60$ $mV$). The lines represent the results of the calculations
using the values of $\Phi_0$ for a CuO$_x$/rubrene interface, known
from our study of the contact resistance: $\Phi_0$ typically varies
between $0.13$ and $0.15$ $eV$ in different devices
\cite{Molinari08}. For the hole effective mass $m_p$, we take the
value extracted from recent optical spectroscopy experiments, which
is close to the free electron mass \cite{Li07}. All other quantities
in the theoretical expression are known, and there are no free
adjustable parameters. As it is apparent from Fig. \ref{FIG. 4}c,
the quantitative agreement between calculations and experimental
data is remarkably good.

This result has several important implications. First, the
comparison shown in Fig. \ref{FIG. 4}c relies on parameters
extracted from completely different experiments, such as
measurements of contact resistance and infrared spectroscopy.
Therefore, the agreement found in our analysis of $\delta V_{Th}(L)$
indicates that our description of the electronic properties of
rubrene is internally consistent at a quantitative level. Second,
our result indicates that describing the low energy density of
states in the valence band of rubrene in terms of the "textbook"
expression for conventional inorganic semiconductors is a good
approximation. Third, and more in general, this work shows how
"simple" measurements of FET electrical characteristics can be used
to extract microscopic parameters---e.g., the carrier effective
mass---that are not otherwise easily accessible experimentally.

The analysis of the threshold voltage behavior as a function of
source-drain bias confirms that the shift in $V_T$ is due to space
charge injected from the contact into the semiconductor. In this
case we estimate the amount of charge injected into the
semiconductor as $C_C V_{DS}$, where $C_C$ is the capacitance
between the injecting electrode and the rubrene crystal (this
concept is identical to that used in the description of space-charge
limited current $I-V$ curves \cite{SCLC}). By reasoning analogously
to the case of the $L$-dependence, we then obtain
\begin{equation}\label{VDSShift1}
       \delta V_T(V_{DS})= \frac{C_C}{C_G}V_{DS}=\frac{\epsilon_C}{\epsilon_D}\frac{t_Ct_D}{L^2}V_{DS}
\end{equation}
Where for the capacitance between contact and crystal, we have taken
as a very simple approximation the expression of a parallel plate
capacitance.

Equation \ref{VDSShift1} predicts that the shift of threshold
voltage is linear in source-drain voltage, which is the case for
$V_{DS}>kT/q$ (see Fig. \ref{FIG. 3}), and that the slope of this
linear relation should scale with the separation between source and
drain as $L^{-2}$, as found experimentally (see inset of Fig.
\ref{FIG. 4}c). From the linear relation between $d[\delta
V_T(V_{DS})]/d V_{DS}$ and $L^{-2}$ we extract the value of
$\epsilon_Ct_Ct_D/\epsilon_D$ which is comparable to that obtained
by directly measuring the crystal thickness (deviations of a factor
of $2-3$ are found in different samples, as can be expected given
the crude estimates of the capacitances).

Note that all the effects that we have investigated here scale
inversely to the gate capacitance $C_G$ (see Eq. \ref{LShift} and
\ref{VDSShift1}). This is a typical signature of the so-called
short-channel effects, which are well-known in silicon devices
\cite{MOSFETSMODELS}. What is remarkable in our organic transistors
is that, owing to the low doping concentration, these effects
dominate already at a fairly large channel length ($L \simeq 1-2 \
\mu m$), i.e. length scales which are typical of devices used in
practical applications \cite{Gelinck04}.

In summary, rubrene single-crystal FETs are sufficiently stable and
reproducible to perform systematic investigations of threshold
voltage shift. From the length dependence of this shift, we have
extracted information about the microscopic properties of rubrene,
such as density of states and a quantitative estimate for the
effective mass in the valence band. In conjunction with a variety of
earlier experiments---infrared spectroscopy, and quantitative
studies of bias and temperature-dependent contact resistance---our
results show that our current picture for the understanding of
organic semiconductors does account for many experimental
observations in a way that is internally consistent at a
quantitative level.

We gratefully acknowledge A. S. Molinari for help with the
experiments. AFM also gratefully acknowledges financial support from
the Dutch NWO-VICI program and from the Swiss NCCR MaNEP.


\begin{references}
\bibitem{SchottkyBarrier}In reference \cite{Molinari08}, we demonstrate that the energy level alignment at a Cu/CuO$_x$/rubrene interface essentially follows the Schottky-Mott limit. This is due
to the presence of the CuO$_x$ tunnel barrier, which causes a weak
coupling between the metal and molecules, resulting in a small
density of interfacial states (5 10$^{11}$ - 1 10$^{12}$ $eV^{-1}
cm^{-2}$).
\end{references}
\end{document}